\documentclass[aps,a4paper,twocolumn,superscriptaddress,preprint,showpacs,showkeys,10pt]{revtex4-1}
\usepackage{graphicx}
\usepackage{subfigure}
\usepackage{latexsym}   

\begin{document}

\title{Microcanonical thermostatistics of coarse-grained proteins with    
			    amyloidogenic propensity}
							
\author{\firstname{Rafael} B. \surname{Frigori}}
\altaffiliation[Present address: ]
   {Universidade Tecnol\'ogica Federal do Paran\'a, Toledo, PR, Brazil}					
\email{frigori@utfpr.edu.br} 

\author{\firstname{Leandro} G. \surname{Rizzi}} %
\email{lerizzi@usp.br} 

\author{\firstname{Nelson} A. \surname{Alves}}
\email{alves@ffclrp.usp.br}
\affiliation{Departamento de F\'{\i}sica, FFCLRP, 
             Universidade de S\~ao Paulo,
             Avenida Bandeirantes, 3900.
             14040-901, Ribeir\~ao Preto, SP, Brazil.}
\date{\today}


\begin{abstract}
   The formation of fibrillar aggregates seems to be a common characteristic of polypeptide chains,
although the observation of these aggregates may depend on appropriate experimental conditions.
	 Partially folded intermediates seem to have an important role in the generation of protein aggregates,
and a mechanism for this fibril formation considers that these intermediates also correspond to metastable states 
with respect to the fibrillar ones.
   Here, using a coarse-grained (CG) off-lattice model,	we carry out a comparative analysis of the thermodynamic 
aspects characterizing the folding transition with respect to the propensity for aggregation of four different systems: 
two isoforms of the amyloid $\beta$-protein, the Src SH3 domain, and the human prion proteins (hPrP).
	 Microcanonical analysis of the data obtained from replica exchange method (REM) is conducted to evaluate 
the free-energy barrier and latent heat in these  models.
	 The simulations of the amyloid $\beta$ isoforms and Src SH3 domain indicated that the folding process 
described by this CG model is related to a negative specific heat, a phenomenon that can only be 
verified in the microcanonical ensemble in first-order phase transitions.
   The CG simulation of the hPrP heteropolymer yielded a continuous folding transition.
   The absence of a free-energy barrier and latent heat favors the presence of partially unfolded 
conformations, and in this context, this thermodynamic aspect could explain the reason why the hPrP heteropolymer 
is more aggregation-prone than the other heteropolymers considered in this study.
   We introduced the hydrophobic radius of gyration as an order parameter and found that it can be used 
to obtain reliable information about the hydrophobic packing and the transition temperatures in the folding process.

\end{abstract}


\maketitle

\section{INTRODUCTION}

   Different diseases, including  Alzheimer's \cite{alzheimer_1,alzheimer_2} (AD), 
Parkinson's \cite{cookson}, and Huntington's \cite{kordower,imarisio} diseases, are known to be the
result of neurodegenerative processes caused by formation of fibrillar aggregates of proteins, 
known as proteinopathies, which occurs primarily in the cellular structures 
\cite{murphy,thirumalai_146,takano,chiti-review,ventura}.
   Because a number of proteins have been found to form fibrillar aggregates without sequence identity or 
structural homology under appropriate conditions in vitro, one can conclude that this is a common 
characteristic of polypeptide chains \cite{bucciantini,thirumalai_146,chiti-review}, that would reflect a 
property of their main backbone.
   Therefore, the use of detailed interatomic potentials in computational experiments employed for the 
thermodynamic characterization of proteins with known aggregation propensity seems to be 
inessential and thus, some general features can be extracted with the aid of simpler models \cite{nussinov}. 
   In particular, we followed ref. \cite{janke-L2006} and utilized a coarse-grained 
off-lattice model \cite{stillinger-2} in our simulations.
   In the simulations, the 20 naturally occurring amino acid residues are replaced with monomers 
whose hydrophobic-polar characters are denoted by $A$ and $B$, respectively.
   Thus, our investigation of the thermodynamic features of peptide chains with known amyloidogenic 
propensity assumes that the nonbonded interactions are described by the hydrophobic-polar 
character of these monomers in all extension.
   The hydrophobicity of the side chains correlates with the aggregation rate 
\cite{dubay,chiti-11}, so it is an important physicochemical ingredient promoting 
the nucleation of fibrillar aggregates.
   The microcanonical analysis of the heteropolymer aggregation process has been performed
\cite{janke-L2006,liang} and supports the main idea that simple models encompassing hydrophobic
interactions among monomers or segments present the overall aggregation phenomenon through a 
temperature-driven first-order transition.
   A more comprehensive study on the properties of the aggregation phenomenon shows that the change 
in the aggregation rate depends on diverse factors, grouped into extrinsic or intrinsic factors 
\cite{dubay,dobson-prone}.
   Extrinsic factors include physicochemical properties related to the polypeptide environment. 
   The intrinsic factors associated with amyloid formation refer to the characteristics
of the polypeptide chains. 
   Factors like hydrophobicity, propensity to form the $\alpha$-helical structure, propensity 
to form the $\beta$-sheet structure, overall charge, and patterns of polar and nonpolar residues
have been demonstrated to influence the aggregation rate.
   The propensity of globular proteins to adopt a different secondary structure content is another
intrinsic factor related to the stability of the polypeptide chain.
   There is strong evidence that partially folded intermediates, such as for the A$\beta$ peptide in the 
case of Alzheimer's, or with substantial unstructured conformations play an important role in the formation 
of amyloid fibril \cite{uversky_131,hamada,neudecker,kumita}.
   Thus, conformational destabilization seems to be a natural requirement for polypeptide chains to achieve 
new arrangements even under physiological conditions, which culminates in their assembly of amyloid fibrils 
\cite{straub-2012}.
	 On the other hand, unstructured or destabilized conformations do not seem to be a necessary 
condition to promote aggregation \cite{liemann,soldi}, although results from mutations on 
the amyloid $\beta$-protein (A$\beta$) have shown that the less stable A$\beta$ variants have 
the fastest nucleation kinetics in fibrillization \cite{ni_stability,bitan}. 
     		
   Here, thermodynamic features of the amyloid $\beta$-peptides ($A\beta$40 and $A\beta$42),
the Src SH3 domain, and the human prion protein (hPrP) are compared {\it in silico}.
    Moreover, we try to associate those features with their aggregation propensity levels.
   The amyloid $\beta$-protein is the principal component of the amyloid plaques found in AD patients 
\cite{rauk_review}.
	 It is a small peptide that is predominantly constituted by 40 or 42 residues.
	 The Src SH3 domain in turn, is a globular domain with 56 residues.
	 Although this domain is non-pathogenic, it is expected to behave quite similarly to prion 
proteins during the aggregation transition.
	 Actually, it has been experimentally shown that the SH3 domain aggregates to form amyloid fibrils 
\cite{ventura,fontana}.
   Molecular dynamics simulation of Src SH3 domains also led to aggregation conformations when the 
amino acid residue interactions were modeled by the G\={o} potential \cite{MD_SH3}.
   The prion protein with PDB entry 1HJM  \cite{calzolai} is considered in this study. 
   A large number of observations have led to the conclusion that partial unfolding of the monomeric cellular 
prion protein (PrP$^{\rm C}$) seems to be necessary for the generation of the misfolded intermediates 
during fibrillization of the scrapie isoform  PrP$^{\rm Sc}$  \cite{liemann,daggett-101,yu-prion-2012}.
 
 	 In addition, we emphasize that thermodynamic equilibrium properties are better analyzed in the 
microcanonical ensemble because it leads to the correct entropic characterization \cite{gross-JCP} and 
sets a simple way of determining free-energy barriers.
   In fact, many systems present equilibrium properties in the microcanonical ensemble but do not have 
their equivalents in the canonical ensemble \cite{barre-PRL2001,bouchet-JSP,costeniuc-PRE}.
   This is a consequence of the nonconcavity of the entropy function, which produces energy-dependent
equilibrium states in the microcanonical ensemble that cannot be associated with any temperature-dependent
equilibrium states in the canonical ensemble.
	 These missed states are the nonequilibrium ones in the canonical ensemble, and they generally correspond
to metastable or unstable states in this ensemble \cite{touchette-PhysicaA}.
   The most relevant feature that emerges from the nonconcavity of the entropy is the
occurrence of a negative microcanonical specific heat.
   Hence, it is preferable to perform a microcanonical analysis so that the metastable states that may 
arise as a consequence of the dynamic structural conversion between the native and unfolded states can be 
properly accounted for.
   In this work, we use the replica-exchange method (REM) \cite{hukushima-REM} and ST-WHAM-MUCA 
\cite{rizzi-st-wham-muca} analysis to obtain such microcanonical results and to base our discussions on.

\section{MODEL AND SIMULATION ALGORITHM}

   To perform the simulations, the primary structure of a protein is mapped through
the Roseman hydrophobicity scale \cite{roseman} onto a string of elements $A$ and $B$
located at the $C_{\alpha}$ atoms.
   Table \ref{tab:roseman} includes the PDB codes used in this study and the corresponding AB sequences.
   The interaction for chains with $N$ monomers is described by the following energy function 
\cite{stillinger-2},
\begin{equation}
  E  =\frac{1}{4}\, \sum_{k=1}^{N-2} \left(1-\cos\theta_{k}\right) \, + \,  
   4 \, \sum_{i=1}^{N-2} \sum_{j=i+2}^{N}
	  \left(\frac{1}{r_{ij}^{12}}-\frac{C\left(\sigma_{i},\sigma_{j}\right)}{r_{ij}^{6}}\right).
\end{equation}
   Here, $\theta_{k}$ is the angle between three successive monomers,
and $r_{ij}$ denotes the distance between the monomers $i$ and $j$ in the chain.
   The coupling constant $C\left(\sigma_{i},\sigma_{j}\right)$
in the Lennard-Jones type potential depends on the pairwise hydrophobic details,
\begin{equation}
C\left(\sigma_{i},\sigma_{j}\right)=  \left\{\begin{array}{ll}
   1 ,  & \sigma_{i}=\sigma_{j}=A  \\
 0.5 ,  & \sigma_{i}=\sigma_{j}=B  \\
-0.5 ,  & \sigma_{i}\neq\sigma_{j} \, .  \end{array} \right.        \label{C_Coeffs}
\end{equation}
   The spherical-cap algorithm \cite{janke-E2005} was used in the simulations in order to update 
conformations.
   This algorithm utilizes spherical coordinates for the generation of new positions for each monomer in the chain.
 
   Conformations at different temperatures were obtained by REM.
   Figure \ref{fig:tempset} presents a log plot of the temperatures used in the replica exchange 
simulations for the heteropolymers A$\beta$, Src SH3, and hPrP.
	 The temperature sets were determined by the following protocol: given an initial inverse temperature 
$\beta_{n}$, we determined the next inverse temperature $\beta_{n+1}$ by simulating only two replicas of 
the system with the Metropolis algorithm and the AB force field.
   First of all, we performed REM using $N_{s}$ sweeps and  $N_{swaps}$ exchange conformation moves 
so as to equilibrate both replicas at the same reference temperature $T_{n}=1/\beta_{n}$, with the Boltzmann 
constant $k_B =1$.
   Next, the inverse temperature of one replica was increased by a small variation 
$\delta \beta$, and $N_{swaps} \times N_{s}$ updates were performed again trying to exchange replicas 
after $N_{s}$ sweeps.
	 If the fraction of accepted replica exchanges $f_{acc}$ was approximately equal to a probability 
$p_{acc}$, this new inverse temperature $\beta_{n+1}$ was accepted as reference; otherwise, 
the inverse temperature was increased by $\delta \beta$ once again.
	 This procedure allowed for the recursive determination of all the temperatures.
	 To compute $f_{acc}$, $N_{swaps} = 50$ and $N_{s} = 2000$ were employed.
	 The acceptance probability $p_{acc}$ was set to $0.40$ for the amyloid $\beta$-peptides, and 
$0.30$ for the Src SH3 domain and the hPrP protein.
	 It has been demonstrated that these $p_{acc}$ values provide a convenient number of round trips  
between extremal temperatures \cite{lingenheil2009,fiore2011}.
	 For the all systems, evaluation of the temperature sets started at the inverse temperature 
$\beta_{1} = 0.5$, with $\delta\beta = 0.01$.
	 A simple statistics with five independent simulations was used for $f_{acc}$ estimation, and thus for
establishment of the set of temperatures $\{T_i\}$ to perform REM simulations for each heteropolymer.

   Final data production was obtained with 12 replicas for A$\beta40$ and A$\beta42$,
16 replicas for Src SH3, and 20 replicas for hPrP.
   Simulations for data production were accomplished with $N_s = 2000$ sweeps and $N_{swaps}=10500$ 
replica exchange moves, but for hPrP the number of exchange moves was doubled.
   Moreover, the above statistics was repeated five times, always starting from different initial conditions.
   Figure \ref{fig:conformations} shows conformations obtained from the AB model for 
the sequences in Table I, and sampled in the transition region.
	
   Data analysis was performed using the ST-WHAM procedure, which is an iteration-free approach that solves
the WHAM equations in terms of intensive variables \cite{st-wham}.
   This procedure is a convenient way of obtaining the microcanonical inverse temperature estimates. 
	 On the basis of the assumption that the density of states can be obtained from
\begin{equation}
    \Omega(E) \propto \frac{H(E)}{W(E)},
\end{equation}
with $H(E)$ being the energy histogram and $W(E)$ the simulation weight, ST-WHAM states that the inverse temperature 
can be estimated from $M$ independent simulations ($n=1, \cdots, M$),
\begin{eqnarray}
  \beta(E) & = &\frac{\partial\, {\rm ln}\, \Omega(E)}{\partial E} \label{eq:st-wham2} \\
               & \approx & \sum_{n} f_{n}^*(\beta_{n}^H + \beta_{n}^W), \label{eq:st-wham3}
\end{eqnarray}
where $f_{n}^* =H_{n}/\sum_{n}H_{n}$, $\beta_{n}^H = \partial\, {\rm ln} H_{n}/{\partial E}$,
and $\beta_{n}^W= -\partial\, {\rm ln} W_{n}/{\partial E}$. 
    The sum in $n$ is over data produced by REM simulations, with the energy histograms 
$H_{n}$ and $W_{n}=e^{-E/T_{n}}$
specified for each temperature $T_{n}$.
    Thermodynamic quantities such as microcanonical entropy $S(E)$ and specific heat $C_v(E)$
are evaluated from a multicanonical entropy-like solution \cite{rizzi-st-wham-muca}, the 
so-called ST-WHAM-MUCA procedure.

\section{NUMERICAL SIMULATIONS AND RESULTS}

	 For comparative purposes, thermodynamic quantities such as the entropy, changes in the 
free energy $\Delta F$, and latent heat $\ell$ are given as a function of the specific energy 
$\varepsilon = E/N$, where $N$ stands for the total number of monomers in the system.
   These quantities are summarized in Table \ref{tab:proteins}.

   Because the force field treats the hydrophobic bonding interactions as the driving force 
in the folding process, it is important to analyze how the hydrophobic monomers behave as a 
function of the temperature.
   In fact, a recent study has shown that, besides hydrogen bonding, hydrophobicity is a fundamental 
interaction in the competing processes leading to folded or misfolded proteins \cite{fitzpatrick}.
   As we will show, the folding behavior can also be investigated by taking into account the
spatial distribution of hydrophobic monomers by means of a radius of gyration restricted only to
this type of monomer.
	 It has been demonstrated that this hydrophobic radius $r_h$ is a suitable scoring function
for discrimination between the native structure and other conformations \cite{alves-rh}.

\subsection{Microcanonical analysis of the A$\beta$ models}

	 Figure \ref{fig:ab40}(a) exhibits the estimates of $\beta(\varepsilon)$ for the 
A$\beta40$ heteropolymer.
   The microcanonical entropy $S(\varepsilon)$ is calculated using these ST-WHAM estimates of 
$\beta(\varepsilon)$ according to the updating procedure ST-WHAM-MUCA.
   This entropy presents the so-called convex intruder (figure not shown) and produces a 
negative specific heat in the energy range 
$[-0.15, -0.07]$,  which is a signature of a first-order phase transition.
  Figure \ref{fig:ab40}(b) depicts the behavior of $C_v(\varepsilon)$.
	A small positive peak in $C_v(\varepsilon)$ is observed at $\varepsilon = -0.579$, 
presumably due to further compaction of the heteropolymer chain. 
  This latter transition is not related to a decrease in the microcanonical entropy with rising energy
in this region.
   Therefore, $\beta(\varepsilon)$ does not present a van der Waals-like curve in 
Fig. \ref{fig:ab40}(b), which results in a continuous phase transition at $\beta_c = 2.695$.

	To evaluate the free-energy profile as the heteropolymer assumes a folded conformation
characterized by a stable phase within the metastable one, we considered the change of the 
microcanonical entropy between these phases in the vicinity of the transition temperature 
$T_f= 1/\beta_f$.
   The energy region of interest is defined by the Maxwell construction in the caloric 
curve $\beta(\varepsilon)$.
	 This region is limited by the energies $\varepsilon_a$ and  $\varepsilon_b$ 
($\varepsilon_a < \varepsilon_b$) where the horizontal straight line of the Maxwell 
construction intercepts the caloric curve $\beta(\varepsilon)$ and identifies the inverse of 
the canonical first-order transition temperature $\beta_{f}$, which is related to the change in the entropy, 
\begin{equation}
 \beta_{f}= \frac{1}{N}\frac{S(\varepsilon_b) - S(\varepsilon_a)}{\varepsilon_b - \varepsilon_a} .
\end{equation}
   The energies $\varepsilon_a$ and $\varepsilon_b$ define the discontinuity or the latent heat, observed
in van der Waals loops.
	 To calculate the change in the free energy $\Delta F(\varepsilon)$ and $\beta_{f}$ 
from $S(\varepsilon)$, we defined a shifted entropy 
between $\varepsilon_a$ and $\varepsilon_b$, 
$\Delta S(\varepsilon) = N( A + \beta_{f} \varepsilon) - S(\varepsilon)$,
where $A$ and $\beta_{f}$ are such that a {\it canonical} entropy,
$S_{can}(\varepsilon) =N( A + \beta_{f}\, \varepsilon)$, is defined in that energy range.
   The constants $A$ and $\beta_{f}$ are such that 
      	$S_{can}(\varepsilon_a)= S(\varepsilon_a)$ and
	      $S_{can}(\varepsilon_b)= S(\varepsilon_b)$.
	These conditions yield $\varepsilon_a = -0.191$ and $\varepsilon_b = -0.026$, 
at $T_{f} = 0.692(1)$ for A$\beta$40.
	
  The corresponding change in the free energy can be estimated as
\begin{equation}
\beta_{f} [F(\varepsilon) -F_a] = \beta_{f}\, N\varepsilon - S(\varepsilon) ,
\end{equation}
where $ \beta_{f}\, F_a=  \beta_{f}\, N\varepsilon_a -S(\varepsilon_a)$ is the reference 
free energy.
   Our results for $\Delta F(\varepsilon)$ are shown in the inset of  Fig. \ref{fig:ab40}(a).
	 This analysis yields the free-energy barrier $\Delta F = 0.038(2)$ at the 
folding temperature $T_{f}$.
   Latent heat is a consequence of the free-energy barrier that prevents the 
system from moving to a stable conformation in the new phase.
   Therefore, the smaller the latent heat, the higher the probability that a spontaneous fluctuation
will give rise to this new phase.
   The estimate for the latent heat per monomer associated with this transition is $\ell=0.165(3)$.

	
	A similar analysis follows for the heteropolymer obtained with the PDB entry 1Z0Q for the 
A$\beta$42 peptide.
  Results are presented in Fig. \ref{fig:ab42}.
	However, for this heteropolymer, the caloric curve displays a less pronounced  van der Waals-like 
behavior around $\varepsilon = -0.2$ as a consequence of a minor nonconcavity of the microcanonical entropy.
  This model yields $\Delta F = 0.014(1)$ (inset of Fig. \ref{fig:ab42}(a)) and latent heat $\ell=0.125(2)$.
  Therefore, the free-energy barrier separating the folded and unfolded states of the heteropolymer 
describing A$\beta42$ is smaller compared with the A$\beta40$ heteropolymer.
  This corresponds to a less severe restriction to possible movements returning the heteropolymer to intermediate 
conformations, depending on how stable the folded conformations are \cite{thirumalai-Nature}. 
  A small positive peak in the specific heat at $\varepsilon = -0.686$ signals a continuous 
transition ($\beta_c= 2.632$) similar to the one observed for the A$\beta40$ heteropolymer.

   The results of the hydrophobic gyration radius $r_h$ for the A$\beta40$ and A$\beta42$ heteropolymers
as a function of temperature are given in Fig. \ref{fig:radius}.
   The folding transition is expected to be accomplished by fast change in the
spatial distribution of the monomers.
   In particular, this change is expected to be highly sensitive to the hydrophobic monomers, 
as illustrated in Fig. \ref{fig:radius}(a).
	 Thus, we hypothesized that the temperature where the derivative \mbox{$<dr_h/dT>$} hits its maximum 
represents the occurrence of a thermodynamic transition. 
	 These temperatures, denoted by $T_{r}$, are easily identified in Fig. \ref{fig:radius}(b):
$T_r =0.68(1)$ and $T_r=0.76(1)$ for A$\beta40$ and A$\beta42$, respectively.
   These figures also identify the latter compaction transitions at temperatures $0.37(1)$ and $0.38(1)$
for the A$\beta40$ and A$\beta42$, respectively.
   Estimates of $T_{r}$ are in very good agreement with the values $T_f$ obtained via Maxwell construction
for both heteropolymers. 
   This indicates that $r_h$ can be considered a reliable order parameter for depiction of a
hydrophobic profile as a function of temperature.
   This conclusion is also supported by results attained for the Src SH3 and hPrP heteropolymers as 
shown in the following subsections.

\subsection{Microcanonical analysis of the Src SH3 model}

   The SH3 domains have attracted much interest because they represent typical examples of proteins 
that fold via a two-state mechanism.
   It is largely accept that the physical process underlying protein folding in these cases is 
based on the nucleation-condensation scenario \cite{sh3-ding,sh3-borreJMB,sh3-borre,sh3-hubner}.
   In fact, a sharp transition at the folding temperature $T_f$ between the 
 unfolded and folded states has been observed \cite{ sh3-ding,sh3-borreJMB}. 

   Results from simulations for the Src SH3 heteropolymer obtained with the AB model are 
shown in Fig. \ref{fig:sh3}.
	 As expected for such peptide with a clear two-state transition mechanism, a 
van der Waals-like loop can be detected in the caloric curve (Fig. \ref{fig:sh3}(a)).
   Therefore, the microcanonical specific heat presents the typical behavior 
observed in a first-order phase transition (Fig. \ref{fig:sh3}(b)).
   The Maxwell construction leads to $\varepsilon_a=-0.182$ and $\varepsilon_b=0.0079$,
with $T_{f} = 0.658(1)$.
   For this model, we obtained a free-energy barrier $\Delta F = 0.068(2)$ (inset in Fig. \ref{fig:sh3}(a)) 
and a latent heat per monomer $\ell=0.190(8)$.
   Two small positive peaks are seen in $C_v(\varepsilon)$ at $\varepsilon =-0.21$ and $\varepsilon =-0.70$.
   Both transitions are likely related to compactions.	
   The behaviour of $r_h$ and its derivative are included in Fig. \ref{fig:radius}.
   The maximum of $<dr_h/dT>$ occurs at $T_{r}= 0.66(1)$.
	 This feature also signals the existence of a second (minor) peak in the derivative of
$r_h$ at $T_r =0.47(3)$, which is related to the transition observed in Fig. \ref{fig:sh3}(a)
at $\varepsilon =-0.70$.

\subsection{Microcanonical analysis of the hPrP model}

    The next heteropolymer corresponds to the extracellular globular domain hPrP with 104 residues.
    Compared to the previous models, the numerical results do not furnish any convex 
intruder in $S(\varepsilon)$.
		The behavior of $\beta(\varepsilon)$, displayed in Fig. \ref{fig:hprp}(a), does not produce a van der Waals-like curve.
	  The microcanonical analysis yields positive specific heats associated with continuous phase transitions (Fig. \ref{fig:hprp}(b)).
    The maxima of $C_v(\varepsilon)$ occur at $\varepsilon_1 =-0.36$ and $\varepsilon_2 = -0.064$, which 
correspond to the phase transition temperatures $T_1 =0.596(2) $ and $T_2 = 0.697(1)$, respectively.
    The behaviour of $r_h$ and its derivative are also included in Fig. \ref{fig:radius}.
    Here, there is a clear transition at $T_{r}=0.60(1)$
and just a tiny change in the derivative of the hydrophobic radius at $T_{r}=0.32(1)$.
			
    Since our findings for this heteropolymer did not evidence a reduction in the microcanonical entropy 
as we moved toward the unfolded states, we may argue the following.
    Either the free-energy barrier separating the folded and unfolded
states for this heteropolymer is not large enough to be revealed by our potential energy
function because it does not contain interactions that reproduce the native contacts, or
in fact the transition does not occur via a two-state mechanism.
    Interestingly, contradictory observations in what concerns 
the prion protein folding have been reported from diverse experiments.
		Experimental observations indicate that the native folding pathway involves
the two-state process despite other evidences to the contrary (see, for example
\cite{yu-prion-2012} and references therein).

\section{Discussion and Conclusions}

   The ability of peptide chains to form fibrillar aggregates seems to follow from their propensity 
to aggregate under conditions that permit partial unfolding.
   These fibrillar structures originate from misfolded protein chains via a complex process.
   A mechanism for the formation of the fibrils considers that the intermediates correspond to metastable states with respect to the fibrillar states.
   In this sense, the existence of these aggregation-prone states can be a consequence of the low degree of stability of the native states.
   It has been demonstrated with a simple lattice model \cite{li-2010} that low-energy conformations populate the aggregation-prone states.
   Such conformations are identified in the variety of lowest energy oligomers and protofilaments obtained when multiple chains are present in the three-dimensional lattice.
   	Here, we performed REM simulations of four biologically inspired heteropolymers to relate the thermodynamic 
properties with their aggregation propensities.
   Considering the amyloid $\beta$-peptides, there is experimental evidence that the A$\beta$42 is more 
aggregation-prone than A$\beta$40 \cite{harper_1997}.
	 Our findings for the free-energy barriers and latent heats of these heteropolymer systems, as listed in 
Table \ref{tab:proteins},	indicate a two-state folding process.
   Smaller $\Delta F$ and $\ell$ for the A$\beta$42 heteropolymer, compared with the respective values
for A$\beta$40, indicate that the formation of native states are facilitated for A$\beta$42.
    On the other hand, these numerical values signal a weaker first-order transition
for A$\beta$42.  
		This may give higher chances for the A$\beta$42 native-like conformations to adopt partially 
folded intermediates under similar stability conditions.
    As a matter of fact, monomeric A$\beta$ peptides are considered to be intrinsically disordered
and therefore, A$\beta$42 can cross the free-energy barrier to misfold conformations
more easily \cite{ni_stability}.

    For the Src SH3 heteropolymer, our results show a stronger first-order phase transition, confirming 
the two-state character expected for this system.
    Thus, on comparative grounds, it is reasonable to understand the experimental requirements to
destabilise the native conformation \cite{liu-src-sh3} to produce some partially unfolded conformations as a
prerequisite for its self-assembly.

	  The conflicting evidence about the mechanism of the folding process involving PrP shows how difficult the experimental measurements are \cite{yu-prion-2012}.
	 Interestingly, our {\it in silico} experiment suggests that if there is a free-energy barrier 
separating the folded and unfolded states of the hPrP heteropolymer, it is not large enough to be revealed 
by our simple potential energy function.
   We argue that the absence of a free-energy barrier favors the presence of partially unfolded conformations 
in the hPrP heteropolymer, which could explain why it is more aggregation-prone than the other heteropolymers.
   Of course, our conclusions are based on a simple force field which was not designed, for example, to reproduce
the native state.
	However, we expect that the hydrophobic force incorporates the main aspects 
that produce the transition-state configurations.
	
   Because the propensity of peptides and proteins to form aggregates depends greatly on the sequence, 
Pawar {\it et al.} \cite{dobson-prone} defined a phenomenological equation to express this feature.  
	 We calculated the intrinsic score $Z_{agg}$ of aggregation for our polypeptide chains at pH 7 by means of the
Zyggregator algorithm \cite{zyggreg}.   
	 This algorithm yielded the values presented in Table \ref{tab:proteins}.
	 Observation of the $Z$-scores reveals that the hPrP polypeptide is more aggregation-prone than the other 
polypeptide chains.
   This conclusion agrees with the thermodynamic phase transition results listed in Table \ref{tab:proteins}
for these different heteropolymer systems, if we consider that weaker phase transitions facilitate the 
coexistence of mixed conformations.
    
	  We also computed the hydrophobic radius of gyration and as can be seen from  Fig. \ref{fig:radius},
it is a convenient order parameter for identification of the folding temperatures.
   The agreement with the microcanonical estimates demonstrates that the spatial packing of hydrophobic 
monomers furnishes reliable information about the folding process.
   More importantly, this quantity can easily be analyzed in any protein folding study and does not
depend on any other information about the protein chain like the usual reaction coordinates needing some
information about native contacts.

\section*{Acknowledgments:}
  The authors acknowledge support by the Brazilian agencies FAPESP, CAPES, and CNPq. R.B.F. was also
supported by UTFPR.
  This work used resources of the LCCA-Laboratory of Advanced Scientific Computation of the University
of S\~ao Paulo.

\newpage
~

\begin{widetext}

\begin{figure}[!th]
\begin{center}
\begin{minipage}[t]{0.5\textwidth}
\includegraphics[width=0.80\textwidth]{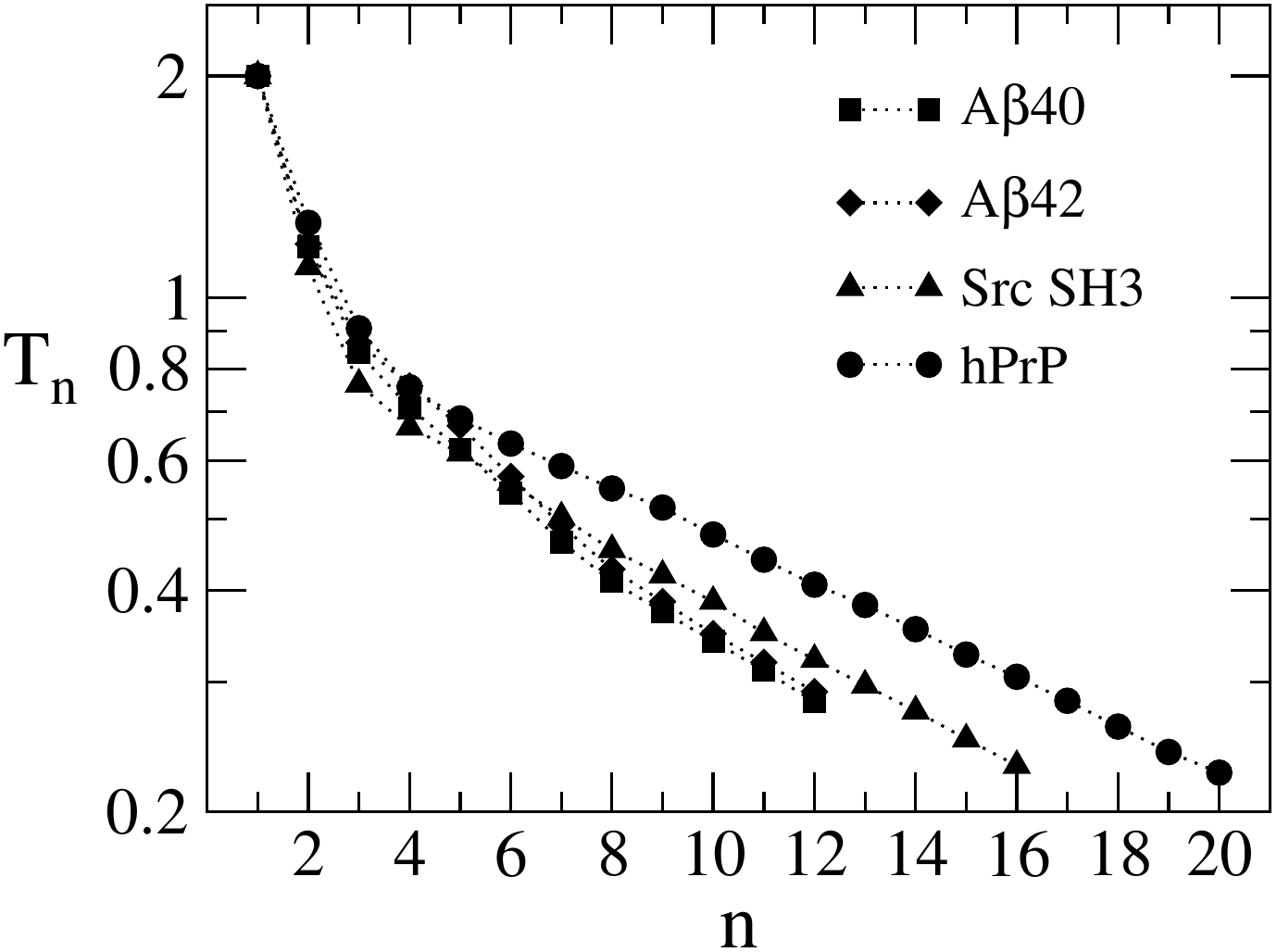}
\end{minipage}
\end{center}
\caption{Temperatures used to perform REM simulations.}
\label{fig:tempset}
\end{figure}

\begin{figure*}[!th]        
\begin{center}
\includegraphics[width=0.80\textwidth]{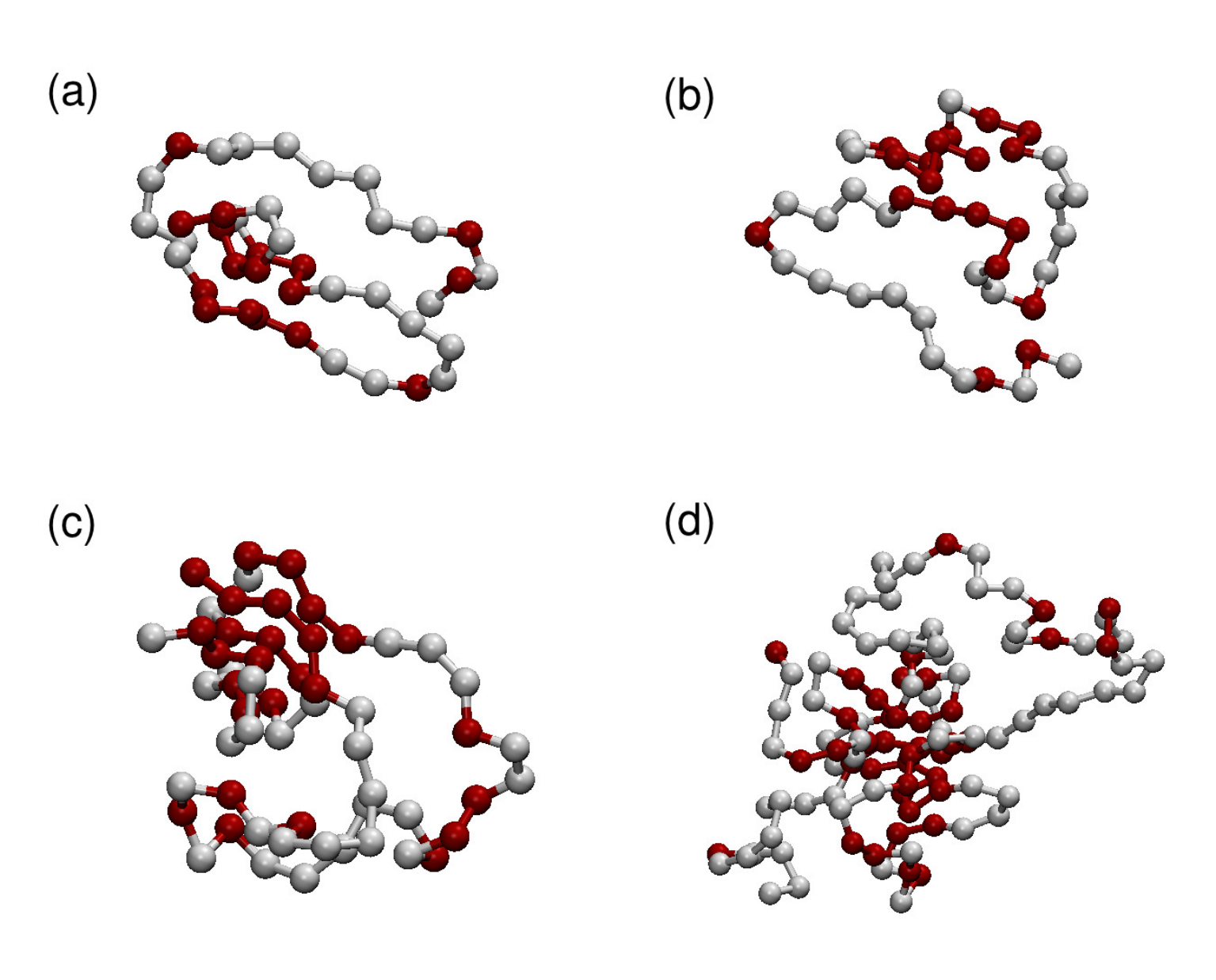}
\end{center}
\caption{Illustrative sampling of conformations for the AB sequences representing 
(a) A$\beta40$, (b) A$\beta42$, (c) Src SH3, and
(d) hPrP proteins in the transition region. Dark (in color: red) color indicates hydrophobic 
monomers. (These images were made with VMD software support).}
\label{fig:conformations}
\end{figure*}

\begin{figure}[!th]
\begin{center}
\begin{minipage}[t]{0.50\textwidth}
\includegraphics[width=0.78\textwidth]{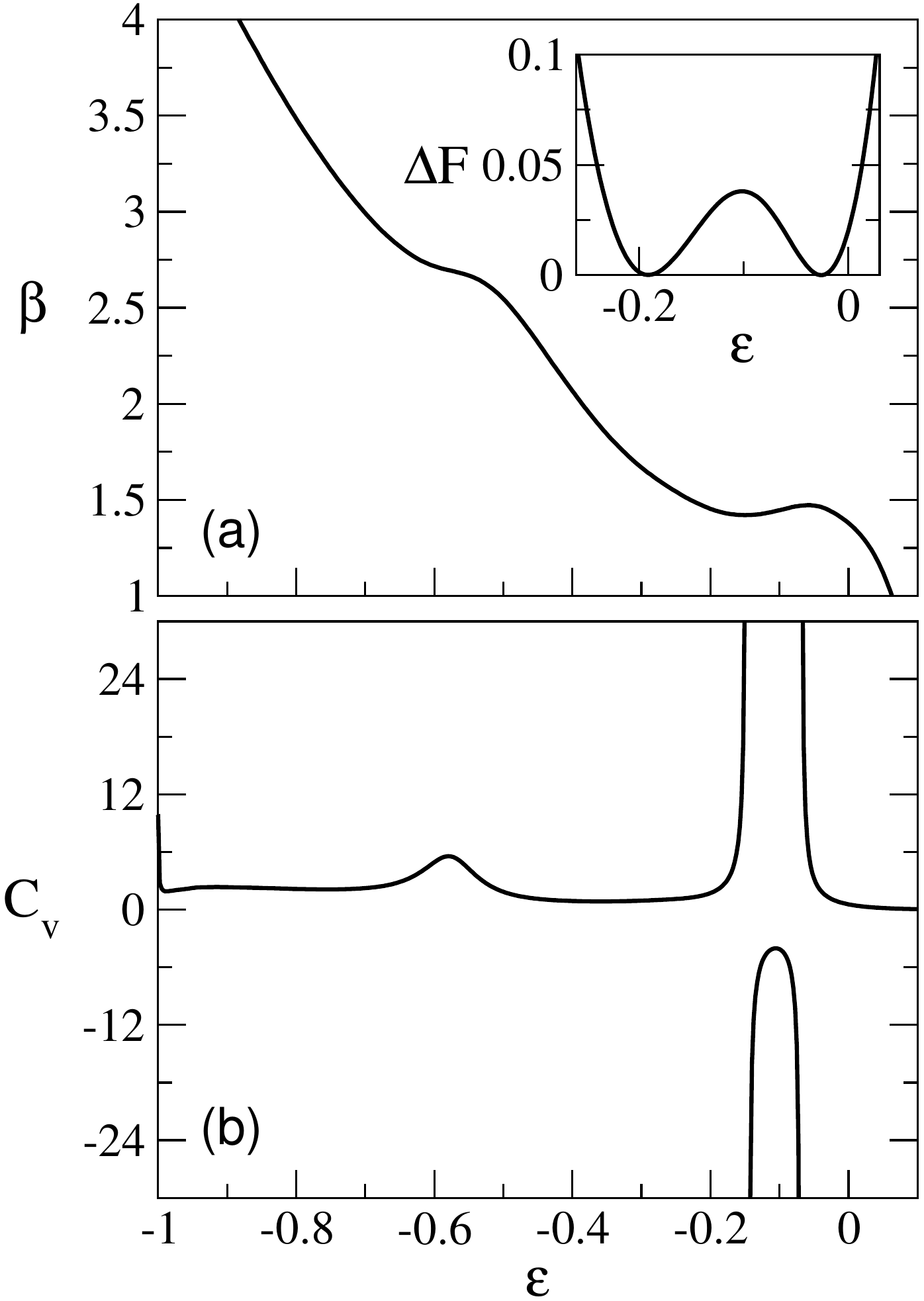}
\end{minipage} 
\end{center}
\caption{
(a) Microcanonical inverse temperature estimates $\beta(\varepsilon) = 1/T(\varepsilon)$, and
(b) microcanonical specific heat $C_v(\varepsilon) = -\beta^2 /(\partial\beta/\partial \varepsilon)$
for the A$\beta40$ heteropolymer. The inset of figure (a) shows the free-energy changes constructed from 
$S(\varepsilon)$ at $\beta_f$.}
\label{fig:ab40}
\end{figure}
	
\begin{figure}[!th]
\begin{center}
\begin{minipage}[t]{0.5\textwidth}
\includegraphics[width=0.78\textwidth]{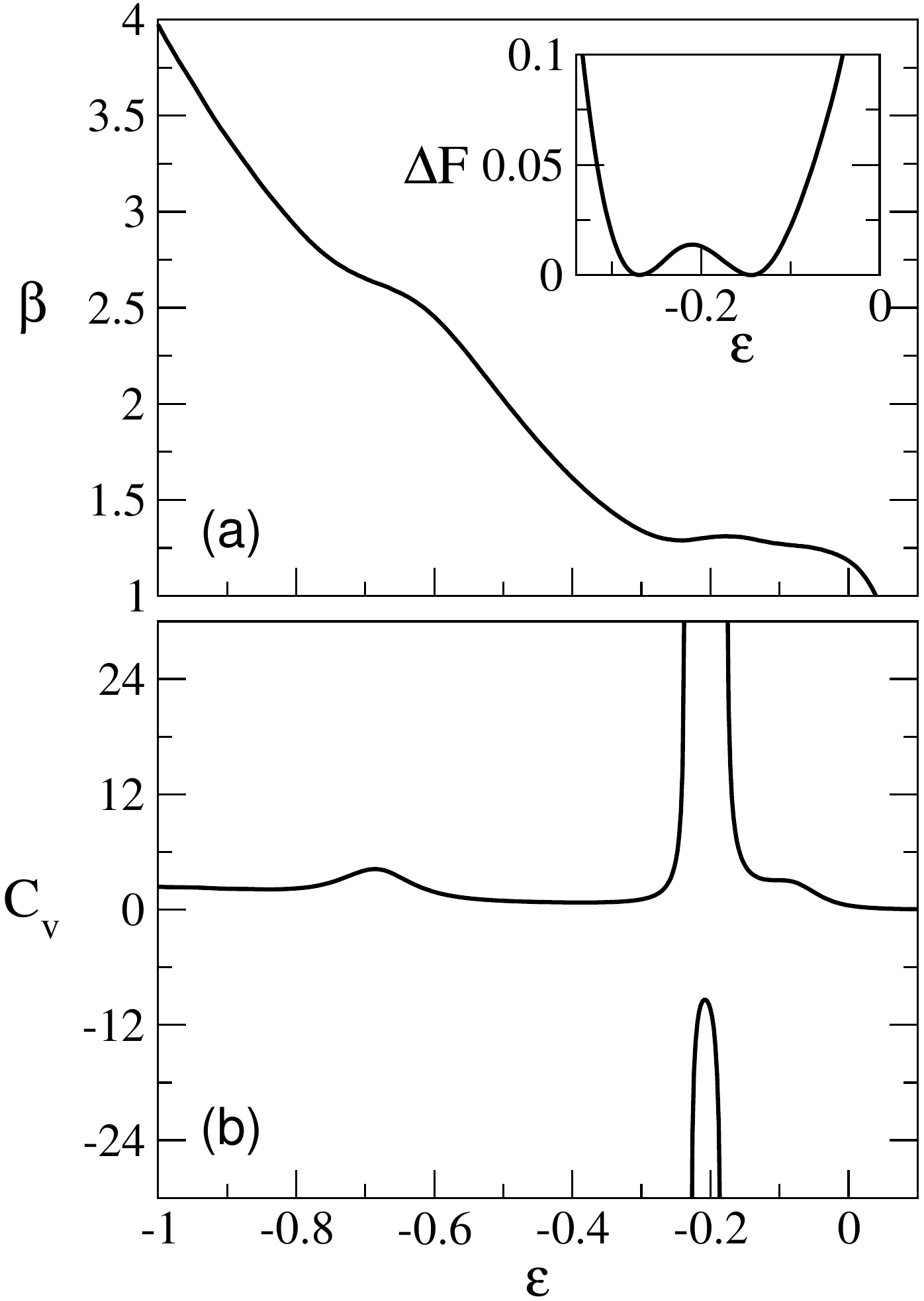}
\end{minipage}
\end{center}
\caption{
(a) Microcanonical inverse temperature estimates $\beta(\varepsilon)$, and
(b) microcanonical specific heat $C_v(\varepsilon)$ for the A$\beta42$ heteropolymer. 
    The inset of figure (a) shows the free-energy changes constructed from 
$S(\varepsilon)$ at $\beta_f$.}
\label{fig:ab42}
\end{figure}
	
\begin{figure}[!th]
\begin{center}
\begin{minipage}{0.5\textwidth}
\includegraphics[width=0.78\textwidth]{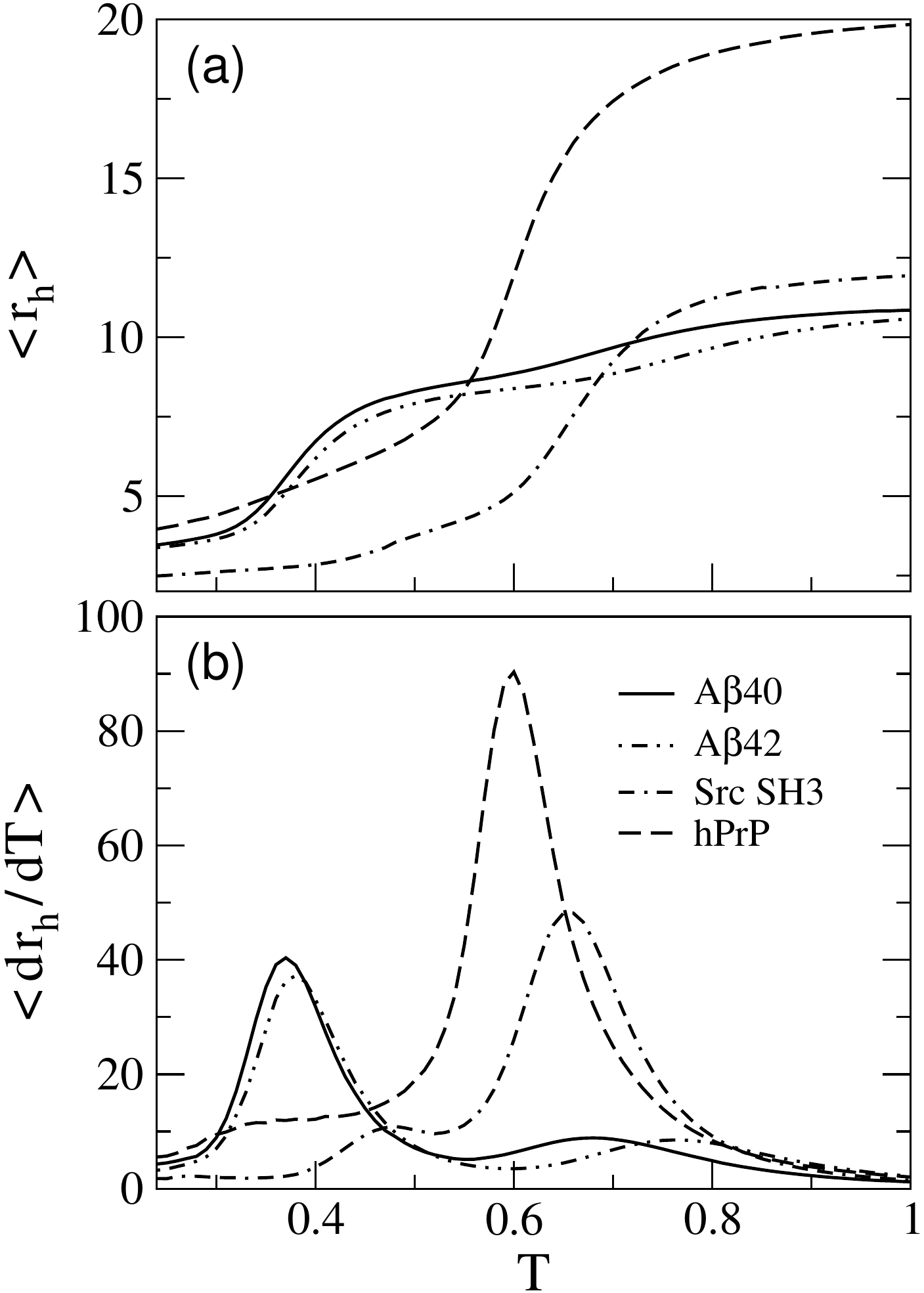}
\end{minipage}
\end{center}
\caption{Behavior of the hydrophobic radius of gyration (a) and its derivative (b) 
         as a function of temperature.}
\label{fig:radius}
\end{figure}

\begin{figure}[!th]
\begin{center}
\begin{minipage}[t]{0.5\textwidth}
\includegraphics[width=0.78\textwidth]{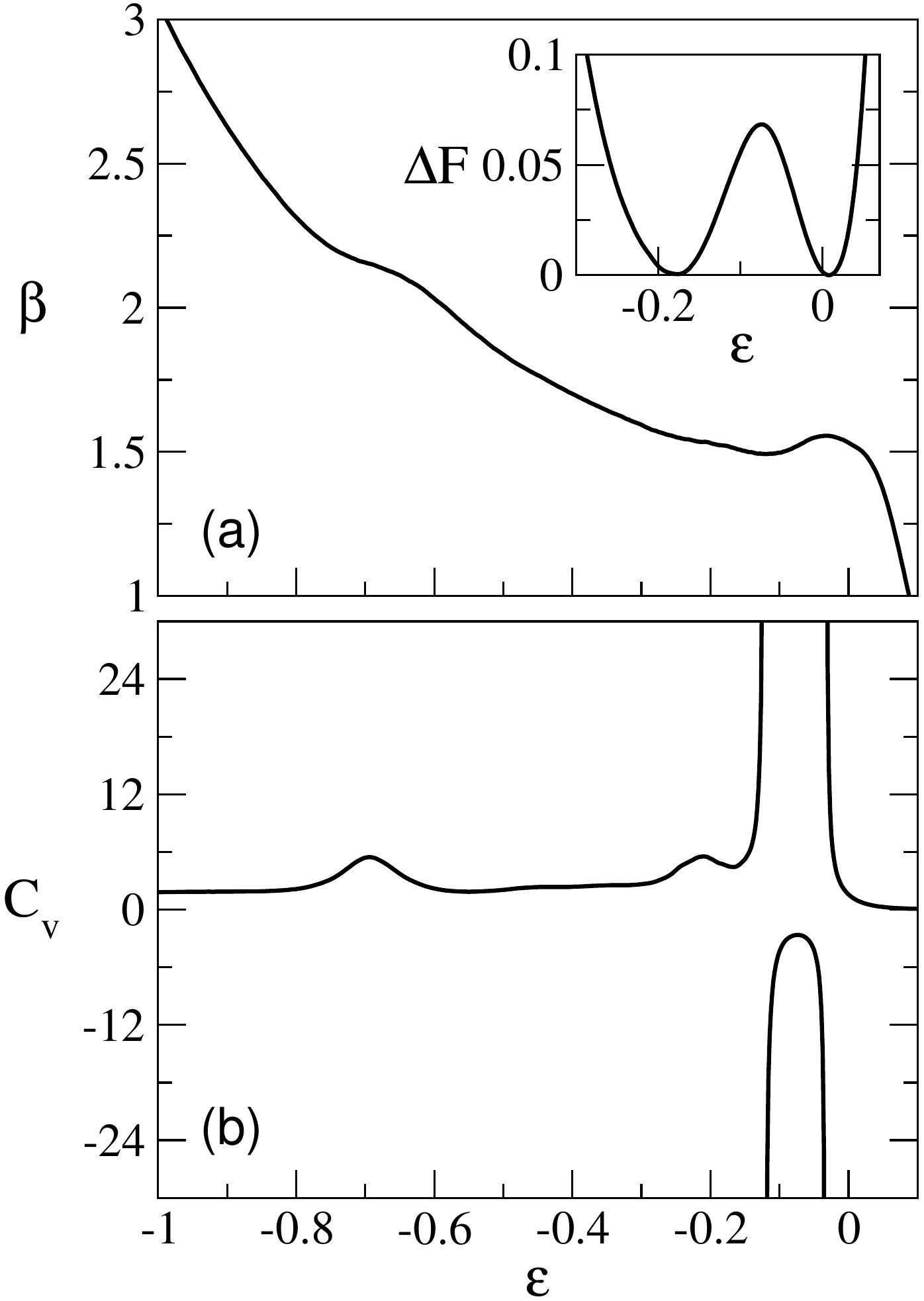}%
\end{minipage}
\end{center}
\caption{ 
(a) Microcanonical inverse temperature estimates $\beta(\varepsilon)$, and
(b) microcanonical specific heat $C_v(\varepsilon)$ for the Src SH3 heteropolymer. 
    The inset of figure (a) shows the free-energy changes constructed from $S(\varepsilon)$ at $\beta_f$.}
\label{fig:sh3}
\end{figure}

\begin{figure}[!th]
\begin{center}
\begin{minipage}[t]{0.5\textwidth}
\includegraphics[width=0.78\textwidth]{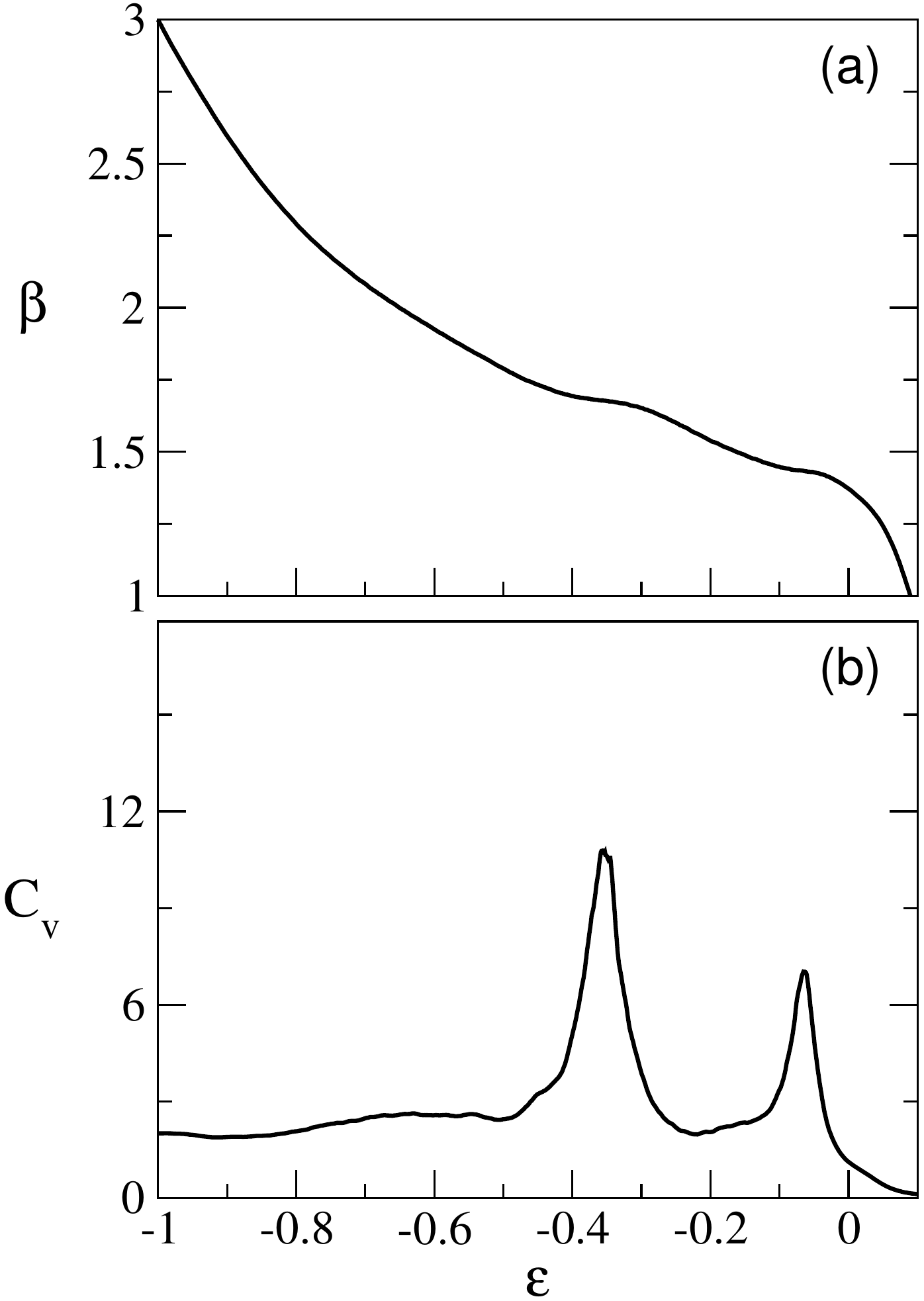}
\end{minipage}
\end{center}
\caption{
(a) Microcanonical inverse temperature estimates $\beta(\varepsilon)$, and
(b) microcanonical specific heat $C_v(\varepsilon)$ for the hPrP heteropolymer.}
\label{fig:hprp}
\end{figure}

~

\begin{table}[!bh]
\centering
\caption{PDB codes and Roseman mapping.}
\label{tab:roseman}
\begin{tabular}{lccl}
\hline
~Protein      &~~~PDB code &~~Residues & ~~~~~~~~~~~~~~~~  Sequence  \\
[0.1cm]
\hline
amyloid $\beta$ &~~ 2LFM   & ~40     & {\ttfamily BABAB~BBBBB~BABBB~BAAAA~ABBAB~BBBBA~AABAA~ABBAA} \\
amyloid $\beta$ &~~ 1Z0Q   & ~42     & {\ttfamily BABAB~BBBBB~BABBB~BAAAA~ABBAB~BBBBA~AABAA~ABBAA~AA}\\
Src SH3	        &~~ 1NLO   & ~56     & {\ttfamily AAAAA~BBBBB~BABAB~ABABB~BBBAB~AABBA~BBBAA~AABBA~AABBA~BBAAB} \\
                &          &         & {\ttfamily BBAAA~B} \\ 
prion protein   &~~ 1HJM   & 104     & {\ttfamily ABBBA~ABBAA~BBAAA~BABBB~BBBBB~BBBBA~BBBAB~BABBB~AABBB~BBBBB} \\
                &          &         & {\ttfamily AABBA~ABAAA~BBBAA~AAAAB~BBBAA~BABAB~AABBA~ABBAA~AABBB~BBBBA} \\
								&          &         & {\ttfamily BBBB} \\
[0.1cm]
\hline
\end{tabular}
\end{table}

\begin{table}[!bh]
\centering
\caption{Comparative results for the heteropolymer models.}
\label{tab:proteins}
\begin{tabular}{cccccc}
\hline
Model  &~ $T_f$    &~ $\Delta F$ &~ $\ell$   &~ $T_{r}$   &~$Z_{agg}$ score \cite{zyggreg}\\
[0.1cm]
\hline
 A$\beta$40  &~0.692(1)  &~ 0.038(2)   &~ 0.165(3) &~ 0.68(1)   &~ 0.90 \\
 A$\beta$42  &~0.769(1)  &~ 0.014(1)   &~ 0.125(2) &~ 0.76(1)   &~ 0.94 \\
 Src SH3     &~0.658(1)  &~ 0.068(2)   &~ 0.190(8) &~ 0.66(1)   &~ 0.96 \\ 
 hPrP        &~0.596(2)  &             &           &~ 0.60(1)   &~ 1.14 \\
[0.1cm]
\hline
\end{tabular}
\end{table}
\end{widetext}

\end{document}